\documentclass[preprint]{aastex63}

\shorttitle{M I + NS binary system 4U~1954+31}
\shortauthors{Hinkle et al.}

\begin{document}

\title{THE M SUPERGIANT HIGH MASS X-RAY BINARY 4U~1954+31}

\author
[0000-0002-2726-4247]
{KENNETH H. HINKLE}
\affil{NSF's National Optical-Infrared Astronomy Research Laboratory,\\
P.O. Box 26732, Tucson, AZ 85726, USA}
\email{hinkle@noao.edu}

\author
[0000-0002-0702-7551]
{THOMAS LEBZELTER}
\affil{Department of Astrophysics, University of Vienna,
T{\"u}rkenschanzstrasse 17, 1180 Vienna, Austria}

\author
[0000-0002-9413-3896]
{FRANCIS C. FEKEL}
\affil{Tennessee State University, Center of Excellence in Information Systems, \\
3500 John A. Merritt Boulevard, Box 9501, Nashville, TN 37209, USA}

\author
[0000-0002-5514-6125]
{OSCAR STRANIERO}
\affil{INAF, Osservatorio Astronomico d'Abruzzo,\\
I-64100 Teramo, Italy, and INFN-LNGS, Assergi (AQ), Italy}

\author
[0000-0003-0201-5241]
{RICHARD R. JOYCE}
\affil{NSF's National Optical-Infrared Astronomy Research Laboratory,\\
P.O. Box 26732, Tucson, AZ 85726, USA}

\author
[0000-0001-7998-226X]
{LISA PRATO}
\affil{Lowell Observatory,
1400 W. Mars Hill Rd, Flagstaff, AZ 86001, USA}

\author
[0000-0003-3682-854X]
{NICOLE KARNATH}
\affil{SOFIA-USRA, NASA Ames Research Center,
MS 232-12, Moffett Field, CA 94035, USA}

\author
[0000-0002-2667-1676]
{NOLAN HABEL}
\affil{University of Toledo, Department of Physics and Astronomy,\\
2801 West Bancroft Street, Toledo, Ohio 43606 USA}

\begin{abstract}
The X-ray binary 4U~1954+31 has been classified as a 
Low Mass X-ray Binary (LMXB) containing a M giant and a neutron
star (NS).  It has also been included 
in the rare class of X-ray 
symbiotic binaries (SyXB).  The Gaia parallax, infrared colors,
spectral type, abundances, and orbital properties of the M star demonstrate
that the cool star in this system is not a low mass giant but a
high mass M supergiant.  Thus, 4U~1954+31 is a High Mass
X-ray Binary (HMXB) containing a late-type supergiant. It is the only known
binary system of this type.  The mass of
the M I is 9$^{+6}_{-2}$ M$_\odot$ giving an age of this system in the range
12 -- 50 Myr with the NS no more than 43 Myr old.  The spin period of
the NS is one of the longest known, 5 hours.  The existence of M I plus NS
binary systems is in accord with stellar evolution theory, with this system a more evolved member 
of the HMXB population. 
\end{abstract}


\section{INTRODUCTION}\label{introduction}

X-ray binaries consist of a neutron star (NS) or black hole (BH) accreting material from a companion star.  
The class of X-ray binaries is divided 
into Low Mass X-ray binaries (LMXB) and High Mass X-ray binaries (HMXB) \citep[see review by][]{tauris_van_den_heuvel_2006}.
The companion star to the NS in the LMXB systems is old, $>$10$^9$ yr, with spectral type 
later than B and mass less than 2 M$_\odot$.  The LMXB are further divided into two classes, low-luminosity dwarf -- NS systems 
and a rare group of late-type giant -- NS systems \citep{liu_et_al_2007}.  The late-type giant -- NS systems
are also classified 
as Symbiotic X-ray binaries \citep[SyXB,][]{murset_et_al_1997}.  
HMXB consist of a massive, young star, with initial mass $\gtrsim$8 M$_\odot$, plus a NS \citep{sander_2018}.  
In HMXB three classes of companion stars are known, main sequence (MS) Be stars, 
supergiant O and B stars, and Roche-lobe filling early-type supergiants \citep{chaty_2011}.   These various 
groups, of course, describe the systems with longer lifetimes or more common 
evolutionary paths and do not cover all the evolutionary products possible \citep{tauris_van_den_heuvel_2006, yungelson_et_al_2019}. 

Symbiotic binaries, labelled SySt, are one of the several varieties of wide binary systems consisting of an evolved star plus a degenerate
object \citep{escorzal_et_al_2020}.  SySt 
consist of a white dwarf (WD) accreting mass from, typically, a K or M giant.   
These objects are characterized by  optical spectra with absorption features of a late-type giant combined with a high excitation 
emission-line spectrum \citep{merrill_1958, kenyon_webbink_1984}.   There is
one late-type star -- NS binary, V2116 Oph, that is generally included in the SySt group.
The SyXB are an unusual subset of the SySt, since most have nearly normal optical spectra. 
SyXB are first identified as X-ray sources and then later associated with M-type optical 
counterparts.  Since the companion star to the NS in this group of LMXB is a giant, the orbital 
periods are measured in years, rather than hours as is the case for LMXB with dwarf companions.   

To date, the total number of confirmed SyXB systems is 
barely over a half dozen with the Galactic population estimated to be $\sim$100--1000 \citep{lu_et_al_2012, yungelson_et_al_2019}.
Two SyXB have well determined orbits, V2116 Oph=GX1+4 \citep{hinkle_et_al_2006} and 
V934 Her=4U~1700+24 \citep{hinkle_et_al_2019}, with orbital periods of 3.2 and 12.0 years, respectively.  The 
stellar companion to the NS is, 
in both cases, a M III with a mass around 1.5 M$_\odot$.  The NS spin in the SyXB binaries is exceedingly 
slow with a period of minutes to 
hours \citep{lu_et_al_2012, enoto_et_al_2014}.  
Radio pulsars all show much shorter spin periods \citep{tauris_van_den_heuvel_2006}.
Very slow pulsars, like those discussed here, are detected in the X-ray only.

To further explore the SyXB class we focus here on 4U~1954+31.  4U~1954+31 is a 
reasonably bright, $V$=10 and $K$=3.5, northern M star.  This star was included in   
a survey of SyXB systems that we are undertaking.  Gaia data revealed it to be overly luminous for a giant.  
We review below multiple lines of evidence that show this M star, the late-type companion of the NS in the
4U~1954+31 binary, to be a massive late-type supergiant.
The supergiant is the evolutionary product of a B star, i.e. the system is a more evolved HMXB. 
The existence of such an object is not 
surprising \citep{nebot_et_al_2015}, although not widely discussed in the X-ray binary literature.
The possibility that the SyXB system Sct X-1 contains a supergiant was discussed by \citet{kaplan_et_al_2007}
but this 
has not been confirmed; Gaia data release 2 does not include a parallax for Sct X-1.
In the following sections we review the literature on 4U~1954+31, derive the parameters for the supergiant, 
analyze the surface abundances, and discuss the evolution of this system.

\section{A BRIEF REVIEW OF 4U~1954+31}\label{history_section}

4U~1954+31\footnote{Some of the X-ray literature uses the alias WISEA J195542.33+320548.8 for 4U~1954+31. 
This is not a recognized alias in Simbad.}  was first detected by $Uhuru$  \citep{forman_et_al_1978}.  
\citet{masetti_et_al_2007}, \citet{corbet_et_al_2008} and \citet{enoto_et_al_2014} review the detections by subsequent surveys.  
The position derived from early X-ray surveys was uncertain and identification with several stars was possible, 
including a Be star \citep{tweedy_et_al_1989}.
$Chandra$ significantly decreased the uncertainty in the X-ray position and resulted in a positional identification 
with a M star \citep{masetti_et_al_2006}.  
The positions are consistent at the 1.8 $\sigma$ level and the identification considered secure.  
From a R ($\lambda/\Delta\lambda$) $\sim$ 800 spectrum in the red the stellar spectral type was found to be a M4--5 III and 
4U~1954+31 classified as an LMXB SyXB \citep{masetti_et_al_2006}.  It appears as a LMXB in the compilation of \citet{bodaghee_et_al_2007}.
A distance of 1.7 kpc was determined by using an absolute magnitude for a M 4 III combined with the observed $V$ and $R$ magnitudes.
\citet{masetti_et_al_2006} note that this is an upper limit to the distance because interstellar reddening had not been taken into account.  

 $Swift/BAT$ data revealed a $\sim$5 hour periodicity. \citet{mattana_et_al_2006} argued that the 5 hr 
 periodicity is incompatible with orbital motion for reasonable NS and M III masses and separations.  Taking the period as the NS spin period, 
4U~1954+31 is one of the slowest known pulsars.  If the NS is rotating at the equilibrium period for disk accretion, the pulsar period implies 
a magnetic field strength of $\sim$10$^{15}$ G \citep{corbet_et_al_2006, corbet_et_al_2008}.
 
Standard masses and the absence of evidence of mass exchange from a contact or CE binary implies a lower limit 
of the orbital period of $\sim$400 days.  
This agrees with the absence of eclipses in the X-ray data.  The large size of the M III and the absence of eclipses 
requires either a highly inclined orbit or a long orbital period \citep{masetti_et_al_2007}.

Several papers also note connections between 4U~1954+31 and HMXB systems. Both the shape of 
the 4U~1954+31 X-ray spectrum and the observed X-ray flaring are in agreement with 
classification as a HMXB \citep{masetti_et_al_2006}.  4U~1954+31 is star number 475 of the INTEGRAL/IBIS
hard X-ray survey and cataloged as a HMXB \citep{krivonos_et_al_2010}.  It also is listed in the BAT all-sky hard X-ray 
survey \citep{baumgartner_et_al_2013} and the IBIS soft gamma-ray 
survey \citep{bird_et_al_2016}. 
\citet{marcu_et_al_2011} measured the NS spin period at 5.3 hr and variable with both spin-up and spin-down which has
parallels in HMXB systems.
\citet{enoto_et_al_2014} discuss extensive $Suzaku$, $Swift/BAT$, and $RXTE/ASM$ observations of 4U~1954+31.  
They found a 5.4 hr NS spin period with $\sim$7\% variations, similar to slowly rotating pulsars
in HMXB.  They conclude that quasi-spherical, subsonic accretion onto the NS is a plausible explanation of the data.  
They assume a $\sim$10$^{13}$ G field for the NS.  
Recurring irregular flares during outbursts with a typical timescale of $\sim$50\,s were 
interpreted as intermittent accretion from the Alfv\'{e}n radius.
\citet{masetti_et_al_2007} and \citet{enoto_et_al_2014} 
detected a narrow Fe-K$\alpha$ line at 6.4 keV, a characteristic of NS in HMXB.

\section{NEW OBSERVATIONS AND REDUCTIONS}\label{observation_section}

Our observational data draw from two sources, a high-resolution near-IR spectrum obtained with 
the Immersion Grating Infrared Spectrometer \citep[IGRINS,][]{park_et_al_2014} at the Lowell Discovery Telescope (LDT) and a time series of 
optical high-resolution spectra taken with the Tennessee State University 2~m Automatic 
Spectroscopic Telescope (AST) and fiber fed echelle spectrograph \citep{ew2007}.

The IGRINS/LDT spectrum was observed on 2018 Nov 23.  The spectrum covers the $H$ and $K$ region at R=45000. 
The reductions follow those of other SyXB spectra observed with IGRINS on the Gemini South Telescope \citep{hinkle_et_al_2019}. The 
initial reduction was done using the IGRINS pipeline. 
The output from this process are the echelle orders which have been ratioed to a telluric reference standard.  
The continuum in each order was removed with a linear fit to the high points.  
The higher order polynomial terms in the continuum  
were removed with the IRAF continuum routine $splot~'t'$ at low order.  
The orders were joined by matching the overlap regions between the orders.  The velocities of the $H$-band CO second 
overtone lines were measured and this velocity is listed in Table
\ref{tab:observations}. 

From 2017 October through 2020 June we observed the spectrum of 4U~1954+31 with the 
Tennessee State University 2 m Automatic Spectroscopic Telescope (AST) and Fiber-Fed Echelle Spectrograph \citep[FFES,][]{ew2007}. 
For these observations the detector was a Fairchild 486 CCD that has a 4096 $\times$ 4096 array 
of 15 $\mu$m pixels \citep{fetal2013}.
The spectra were acquired with a fiber that results in a resolving power of 25000 at 6000~\AA\ and 
contain 48 orders that range from 3800 to 8260~\AA.
Unfortunately, the S/N of these spectra is insufficient for abundance analysis.
The observations are listed in Table \ref{tab:observations} and were used for velocities.  

An overview of the velocity reductions for the AST data can be found in \citet{fetal2009}. 
For 4U~1954+31 a set of 40 lines were used that range in wavelength from 5000~\AA\ to 6800~\AA. 
These lines were selected from a more extensive solar-type star line list based on the line 
being relatively unblended in M-giant spectra. Zero point error in the AST data \citep{fetal2013} was corrected by 
comparing unpublished radial velocities of IAU radial velocity standards, measured from 2~m AST spectra, with 
the mean values of the same stars published in \citet{scarfeetal1990}.  This indicated that the AST data needed a 
velocity correction of 0.6 km~s$^{-1}$ and this 
was applied to each of our AST velocities.

\section{STELLAR AND BINARY SYSTEM PARAMETERS}\label{parameters_section}

\subsection{Effective Temperature, Luminosity, Radius, and Mass}

\begin{figure}
\centering
\includegraphics[width=.8\linewidth]{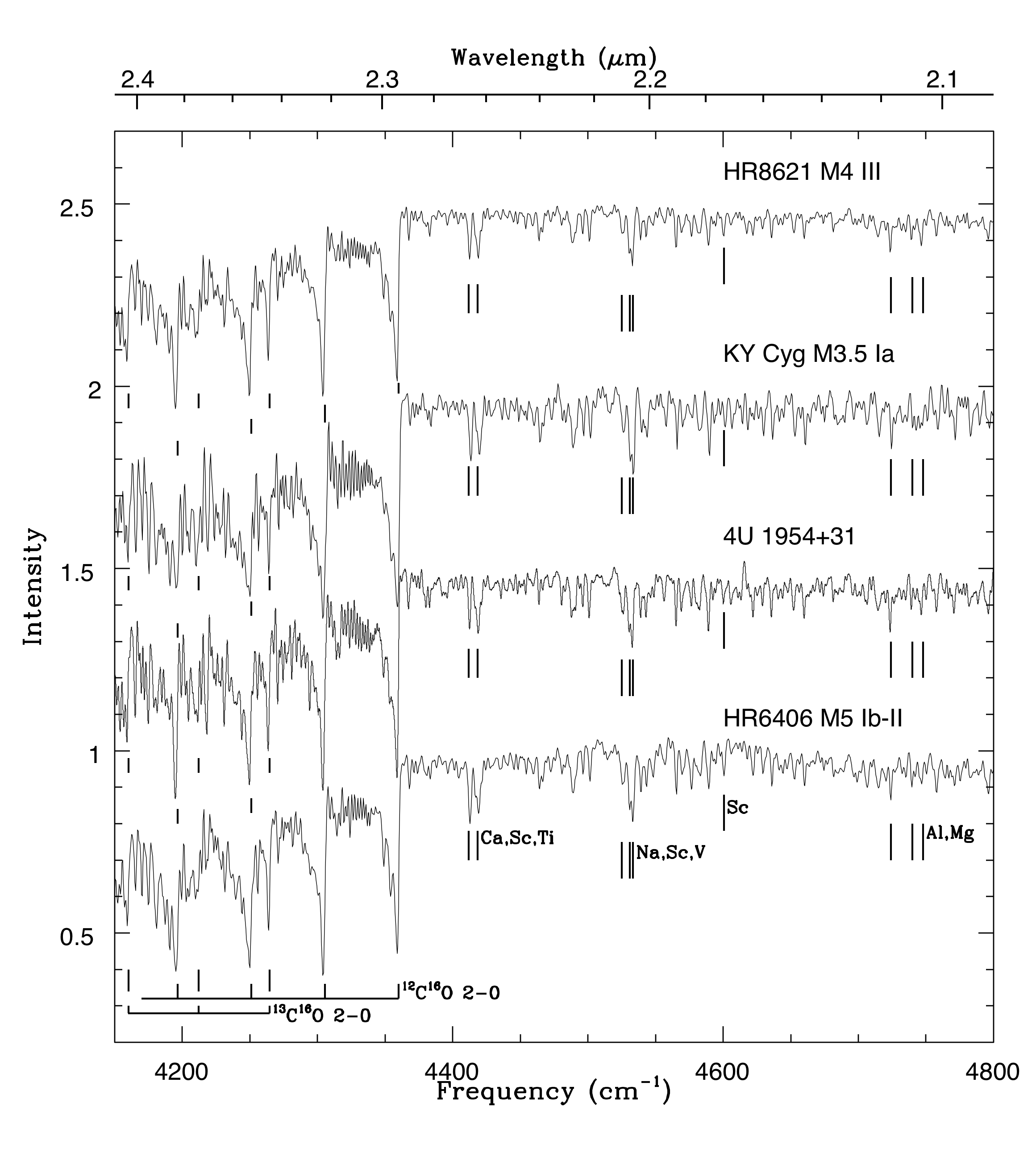}
\caption{
The IGRINS K band spectrum of 4U~1954+31 convolved to R=3000 and
compared to standard star spectra from \citet{wallace_hinkle_1997}.
The molecular and atomic lines in 4U~1954+31 are conspicuously
stronger than in the M4 III spectrum (top) demonstrating that
4U~1954+31 is indeed a M supergiant.  The CO bands are stronger
than in the M5 Ib-II spectrum and similar to the M3.5 Ia spectrum.
The Brackett $\gamma$ emission feature at 4616 cm$^{-1}$ in 4U~1954+31
is a reduction artifact.
}
\label{fig:spec_type}
\end{figure}

\citet{masetti_et_al_2006} found a spectral type for 4U~1954+31 of M4--5 III.
However, comparison of the Gaia \citep{Gai2018} distance, 3295 $^{+985}_{-631}$ pc  \citep{bailer-jones_et_al_2018}, and 
the optical/near-IR magnitudes of 4U~1954+31 implies that the M 
star is much more luminous than a giant. 
The high resolution IGRINS spectrum was convolved to a resolution R = 3000 and compared with K band standard 
star spectra of \citet{wallace_hinkle_1997}.  As shown in Figure\,\ref{fig:spec_type} the spectrum matches 
that of a M4 supergiant and does not match a M4 giant. 

To quantify this result the effective temperature was determined by an analysis of the near-IR spectra.  
Using the technique discussed by \citet{hinkle_et_al_2016}, the H band second overtone vibration-rotation CO lines were measured.  
Over 100  minimally blended, moderately weak lines were identified and the CO excitation temperature, 3200$\pm$100 K, was 
found using curve-of-growth techniques.  From Figure 1 of \citet{lebzelter_et_al_2019} the corresponding effective 
temperature is 3340$\pm$240 K.  The \citet{lebzelter_et_al_2019} calibration of CO excitation temperature 
versus effective temperature employs M III stars. 
The calibration for M supergiants was confirmed by measuring the CO excitation temperature for $\alpha$ Ori and $\alpha$ Her
using archival FTS data from the 4 m FTS \citep{pilachowski_et_al_2017}.  
The 4U~1954+31 CO excitation temperature of 3200 K is bracketed by the 3300 K value for the M2 Ia 
star $\alpha$ Ori and the 3150 K value for the M 5 Ib--II star $\alpha$ Her.  
Using the calibration of \citet{lebzelter_et_al_2019} these convert to effective temperatures 
of 3590 K for alpha Ori, and 3220 K for alpha Her, in agreement with their spectral types (M2 Ia and M5 Ib--II).
The spectral type - effective temperature calibration of \citet{levesque_et_al_2005} gives an effective 
temperature of a M4 I as 3535 K, within our uncertainty.

\begin{figure}
\centering
\includegraphics[width=.8\linewidth]{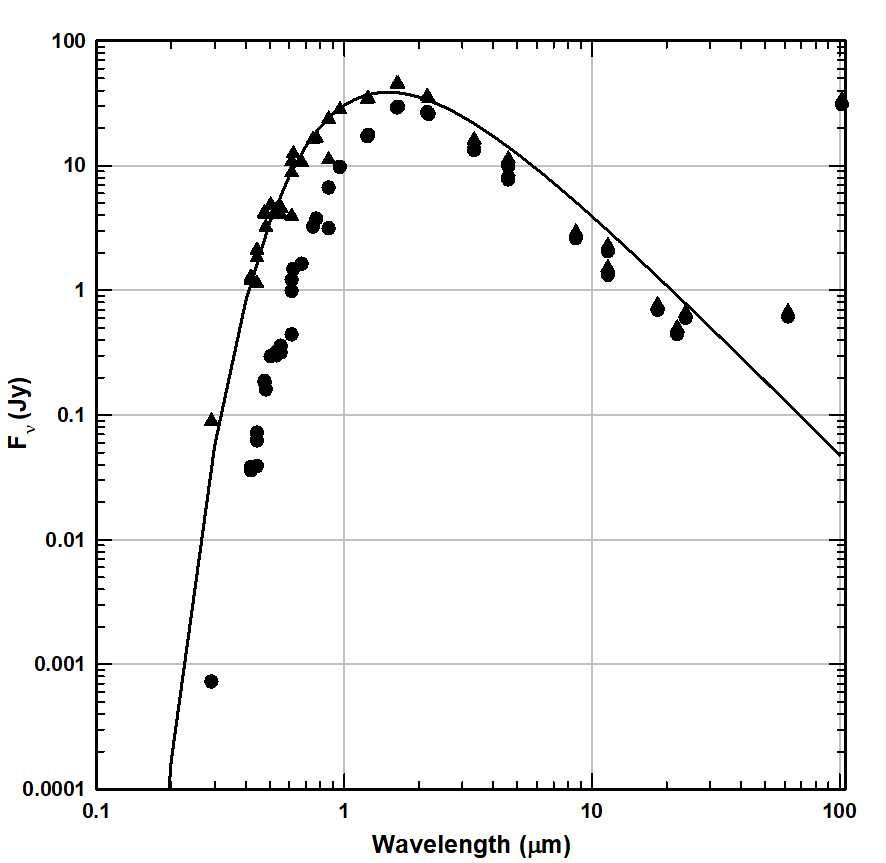}
\caption{The SED for 4U~1954+31.
Circles -- Photometry for 4U~1954+31 taken from the VIZIER data base.
Triangles -- Photometry de-reddened to (V-K)=4.  Line -- 3400 K blackbody. 
}\label{fig:photometry_lit}
\end{figure}

Using literature photometry a spectral energy distribution (SED) for 4U~1954+31 was developed (Figure\,\ref{fig:photometry_lit}).  
From the compilation of \citet{schlafly_finkbeiner_2011} the visual extinction along the line of sight 
toward 4U~1954+31 is A$_v$ = 4.48.  
In Figure\,\ref{fig:photometry_lit} a 3400 K blackbody is compared to the photometry dereddened by (V-K)$_0$=4.0.  
Averaging the various estimates for the effective temperature we adopt 3450$^{+100}_{-50}$ K.
A 3450 K blackbody at the 3295 $^{+985}_{-631}$ pc Gaia distance 
corresponds to a stellar radius of 586 $^{+188}_{-127}$ R$_\odot$.  The luminosity is 43880$^{+34070}_{-15900}$ L$_\odot$.  

\begin{figure}
\centering
\includegraphics[width=.8\linewidth]{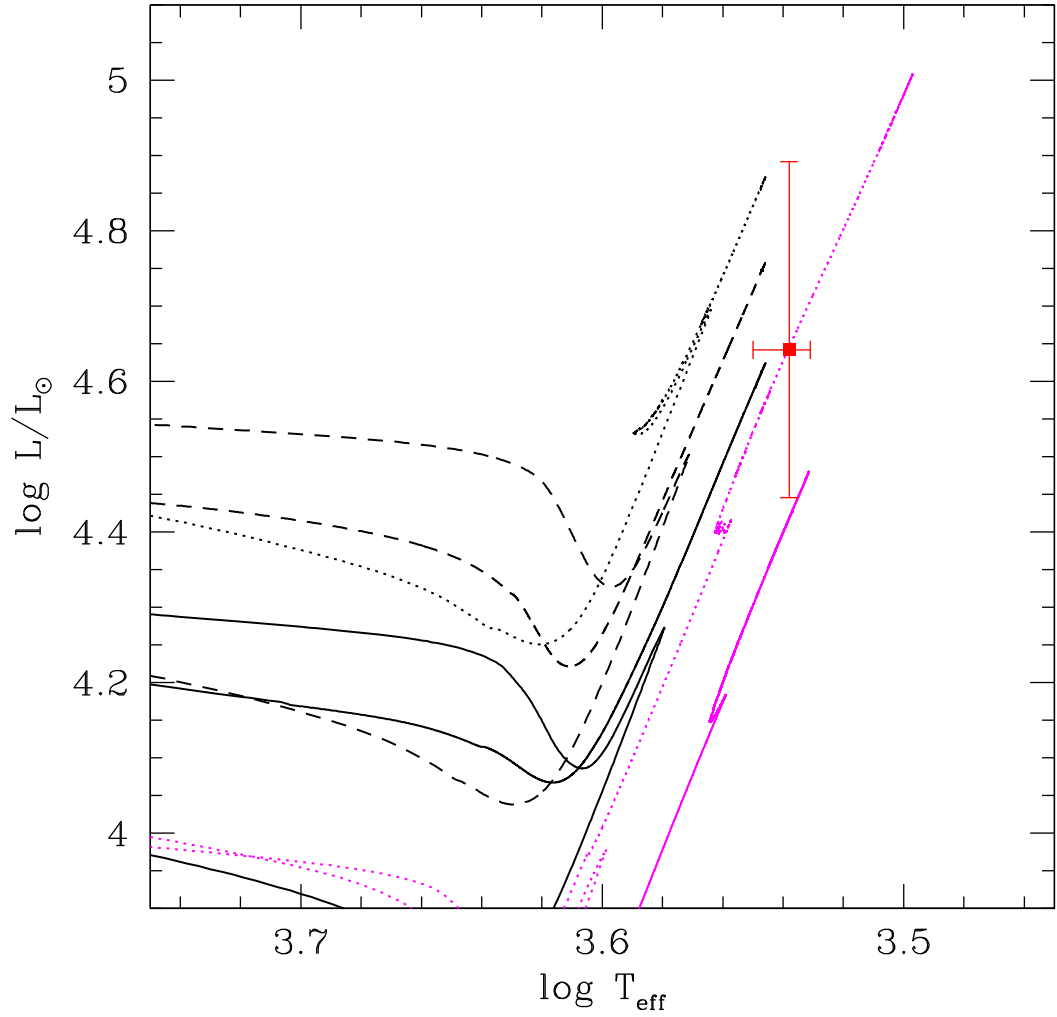}
\caption{Location of the M supergiant component (red square with uncertainties shown) of the 4U~1954+31 binary on the HRD. Lines 
denote FuNS evolutionary tracks for 7\,$M_{\sun}$ (solid magenta), 
9\,$M_{\sun}$ (dotted magenta), 11\,$M_{\sun}$ (solid black), 13\,$M_{\sun}$ (dashed black), 
and 15\,$M_{\sun}$ (dotted black).
For $M\leq 13$ $M_{\sun}$ the evolutionary tracks  loop to higher temperatures,
beyond the 5000 K boundary of the abscissa,
during the core-He burning phase.}
\label{fig:hrd}
\end{figure}

Figure\,\ref{fig:hrd} compares the position of the M4 I component of 4U~1954+31 in 
the HR diagram with standard evolutionary tracks for 7, 9, 11, 13, and 15\,$M_{\sun}$ stars from 
non-interacting and non-rotating FuNS\footnote{Software: FUll Network 
Stellar \citep{straniero_et_al_2006}.} evolutionary models \citep{straniero_et_al_2019}. 
All the tracks in Figure\,\ref{fig:hrd} refer to solar composition models.
As appropriate for red supergiants, advanced 
evolution beyond the H and He burning phase was included.
Accordingly, our tracks with $M\geq 10$ $M_\sun$ follow the classical evolution of massive stars: H, He, C, Ne, O and Si burnings. 
Computations have been stopped after the Si burning phase, at the onset of the full nuclear statistical 
equilibrium (NSE), briefly before the core collapse. On the contrary, after the core-He burning, the 7 and 9 $M_\sun$ models undergo a degenerate C burning, at the end of which they experience a super-AGB phase. In this case, the computations have been  stopped after a few thermal pulses. Note that the luminosity  increases at the beginning of the super-AGB phase because of the occurrence of the hot bottom burning \citep{doherty_et_al_2014}. This phenomenon is stronger in the 9 $M_\sun$ model.  
The luminosity and effective temperature of 4U~1954+31 compared to the evolutionary tracks shows that the most likely mass is 9\,$M_{\sun}$.
Uncertainty brackets masses of 7 to 15 \,$M_{\sun}$.
The corresponding surface gravity is 0.6 - 0.8 cm s$^{-2}$, i.e. log\,g $\sim$ -0.15.  
Table \ref{tab:system_parameters} summarizes the stellar parameters derived for 4U~1954+31.

\subsection{Abundances}\label{abund_section}

Abundances were measured using the spectral synthesis technique applied to the near-infrared IGRINS spectrum. 
As noted in section \ref{observation_section}, the spectra in the visual range were not usable for such an analysis.
Hydrostatic atmospheres from the COMARCS grid of models by \citet{aringer_et_al_2016} 
were employed and the spectra computed with the help of the COMA code \citep[][and 
references therein]{aringer_et_al_2016}. 
For the molecular lines of CO, CN, and OH the line list of \citet{li_et_al_2015} 
was selected with some modifications of line positions based on the Arcturus atlas \citep{hinkle_et_al_1995}, 
\citet{brooke2014}, and HITRAN08\footnote{https://hitran.org}, respectively.
A standard model mass of 10\,$M_{\sun}$ was used for all model calculations.

For the synthesis of the 4U~1954+31 model atmospheres were selected with T$_{eff}$=3400\,K and 3500\,K and log~$g$=0.0 and -0.5 
from the COMARCS grid to match the derived stellar parameters (Table \ref{tab:system_parameters}).
The microturbulence velocity was set to 3\,km\,s$^{-1}$.  This value has been used 
for $\alpha$\,Ori and a variety of other supergiants \citep[e.g.][]{carr_et_al_2000}.
The microturbulence was tested by using a macroturbulence velocity of 4\,km\,s$^{-1}$ to model a few unblended atomic lines.
Molecular lines are hardly affected by the exact value of the microturbulence.
Test calculations were also done altering the microturbulence to 2.5\,km\,s$^{-1}$. The 
small deviation found was included in the uncertainties.
As a starting point a solar composition was assumed. 
Beside the factors mentioned above, the setting of the continuum and line-by-line differences are included in the error budget.
Non-LTE effects have not been taken into account.

The observed spectral region is extensive and practical limitations resulted in fitting sub-sections of 
the spectrum.  Sets of lines for each studied element were used to 
guide the determination of the abundance (Tables \ref{t:usedatomiclines} and \ref{t:usedmollines}). 
For instance, for CO, we combined an overall fit to the $\Delta$v=3 band heads listed 
in Table \ref{t:usedmollines} with fits of individual lines in the H- and K-band.
The line selection took into account the extent and quality of telluric correction at the corresponding 
wavelength, the presence of blending features, the general quality of the model fit in the surrounding 
parts of the spectrum, and in the case of the molecular lines, coverage of a broad range of transition bands and excitation levels. 
Papers by \citet{carr_et_al_2000} and \citet{smith_et_al_2013} were used as a starting point for the line selection.
Identifications and positions were taken from the Arcturus atlas. 

Abundances of the key elements C, N, and O were derived by altering each abundance 
until a simultaneous fit of the lines of CO, CN, and OH was achieved.
The abundances of other elements were determined by choosing for each line the best 
fitting model from a small grid altering the abundances in steps of 0.05 dex. The 
elemental abundance was then computed as the mean from all lines considered.
Changing the model temperature by up to 100 K resulted in a very small change in the line depths
and thus the derived abundances. The abundances were more sensitive to log\,g.  Changing 
log\,g from 0.0 to -0.5 leads to a strengthening of the atomic lines and as a consequence 
a systematic decrease of the abundances by typically 0.1 dex.

As a test, a high-resolution $H$- and $K$-band spectrum of $\alpha$ Ori from 
the Kitt Peak FTS archive was synthesized.  Using a data set extending from the near- into the thermal-IR, 
\citet{lambert_et_al_1984} derived [C/Fe]=$-$0.4, [N/Fe]=+0.6, and [O/Fe]=$-$0.2 
for this star with typical uncertainties of 0.15 dex. 
The  $\alpha$ Ori stellar parameters of \citet{lambert_et_al_1984}, 
T$_{\rm eff}$=3800\,K, log\,g=0.0, and microturbulence $\xi$=3\,km\,s$^{-1}$, were selected for the test.
The \citet{lambert_et_al_1984} carbon and nitrogen abundances were reproduced 
within the error bars although a slightly higher oxygen abundance was 
found.  This likely results from the use by \citet{lambert_et_al_1984} of OH $\Delta$v=1 lines that 
lie outside the $H$ and $K$ windows.

$^{12}$CO lines in the 4U~1954+31 spectrum with central depths exceeding 60\% could not be fit by our model spectra.
This phenomenon is well known for the K-band spectra of highly evolved stars and likely 
indicates limitations of the atmospheric structure of the hydrostatic models \citep{lebzelter_et_al_2019}.
Such lines were not included in our abundance analysis.
We found that CN lines in the H-band give tentatively lower abundances of N than in the K-band. 
Since more isolated lines were accessible in the $K$-band than the H-band, we decided to put a 
higher weight on the results from the $K$-band.  

Table~\ref{tab:abundances} lists abundances obtained from the spectra
for $T_{\rm{eff}}$ = 3400\,K and log~$g$ = 0.0. 4U~1934+31 is slightly metal poor.  The mean 
of the Ti and Fe abundance, elements represented by a good sample of lines, gives [M/H]=$-$0.33.
An overabundance was found for Sc and Mn (average +0.25).
Using a selection of $^{13}$CO lines the $^{12}$C/$^{13}$C ratio was determined by spectrum synthesis to be 8$\pm$2. 
Using the semi-empirical analysis of \citep{hinkle_et_al_2016} the $^{12}$C/$^{13}$C was found to be 9$\pm$2, 
in agreement with the synthesis.
For the oxygen isotopes, $^{16}$O/$^{17}$O = 2000$^{+1000}_{-600}$ and $^{16}$O/$^{18}$O = 1400 (single line).

\subsection{Orbit}\label{orbit_section}

\begin{figure}
\centering
\includegraphics[width=\linewidth]{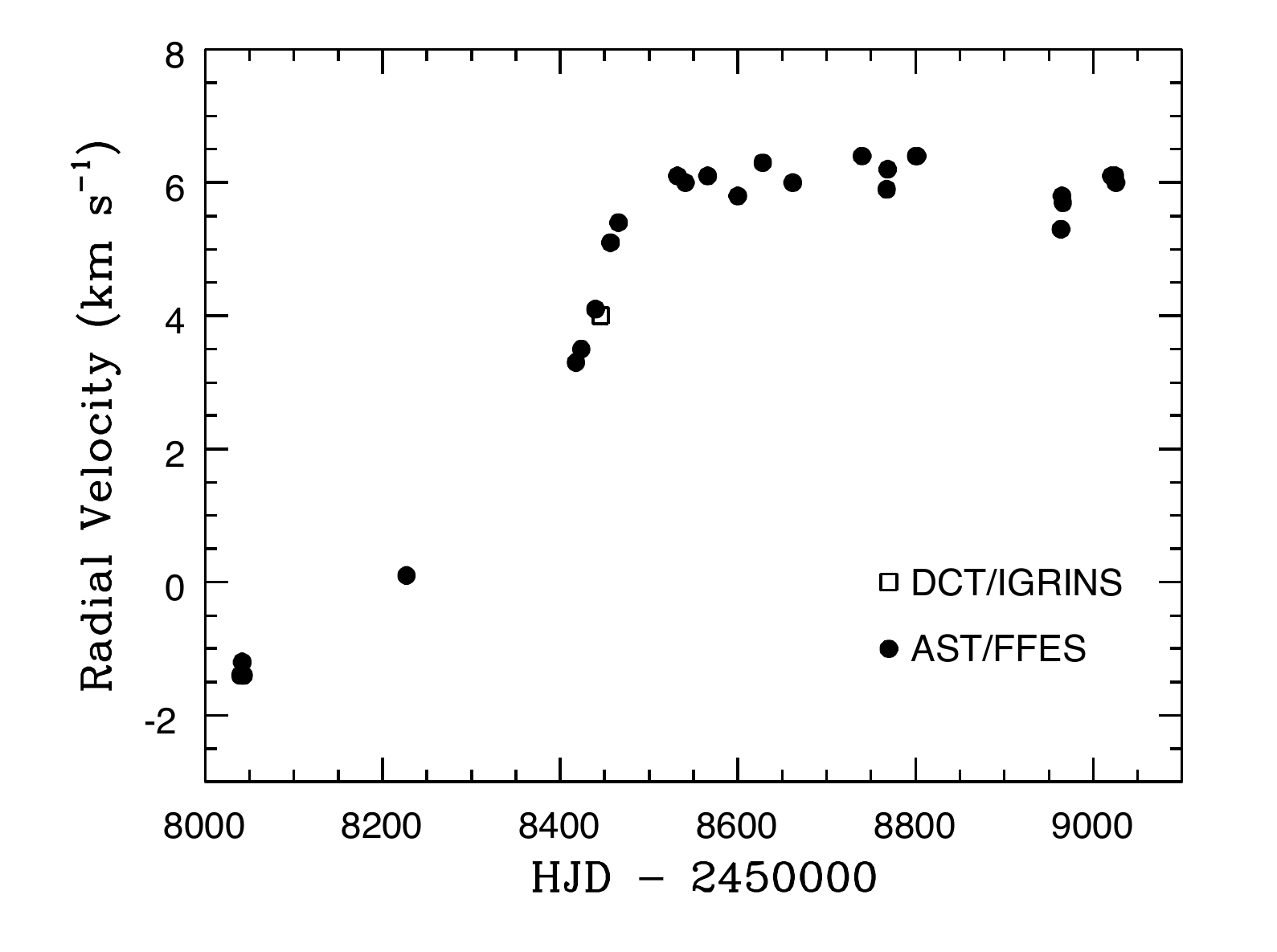}
\caption{
Radial velocities of the M I component of 4U~1954+31 from Table 4.  The velocities span an interval of 2.7 yrs (Table \ref{tab:observations}).  The period of the orbital motion is clearly much larger.
}
\label{fig:orbit}
\end{figure}

The goal of the velocity monitoring was to determine a single-lined spectroscopic orbit that 
would further 
constrain the physical properties of the system\footnote{The convention concerning primary and 
secondary in this system can be confusing since we discuss both initial and final masses.  The terms primary and secondary are be used only
where the meaning is unambiguous.}.  Velocities from AST are shown in Figure \ref{fig:orbit}.  After 2.5 years of 
monitoring, orbital motion is clearly present but the velocities have not yet closed.  From  
the velocities three conclusions are apparent. The period of the orbit is $\gtrsim$3 yrs, the orbit is eccentric, 
and the systemic ($\gamma$) velocity is $\sim$4 km s$^{-1}$.

Systems that fill their
Roche lobes undergo rapid mass transfer and, possibly, a common envelope episode. This 
is clearly not the case for 4U~1954+31, so that the radius of the supergiant
must be smaller than the corresponding Roche lobe. The Roche lobe radius of the M star ($R_L$) is governed by the 
mass ratio, q=$M_{MI}/M_{NS}$,  and related to the binary separation, $a$,  by: 

\begin{equation}
    \frac{R_L}{a}=\frac{0.49q^{\frac{2}{3}}}{0.6q^{\frac{2}{3}}+ln(1+q^{\frac{1}{3}})}
\end{equation}

\citep{eggleton_1983}.  
Assuming a typical neutron star mass of 1.4 M$_\odot$ and a mass of
9 M$_\odot$ for the 
M supergiant, the condition that $R_L$ is larger than the estimated stellar 
radius, i.e., 590\,$R_{\sun}$ (Table \ref{tab:system_parameters}), implies a minimum value 
for the binary separation of $a\,\approx\,1090\,R_{\sun}$ ($\sim$5 AU). Then,
Kepler's third law gives a minimum orbital period of 3.55 yrs.  The current observations limit the orbital period to $\gtrsim$
3 yrs corresponding to a binary separation $\gtrsim$ 4.8 AU, in agreement with the supergiant nature of the M star.  
For instance, with a 3 yr period and the Roche lobe greater than the estimated stellar radius, the supergiant mass should be $\leq 11.6$ M$_\odot$. 
However, a longer period would allow even higher masses.

In principle, the Corbet diagram could provide an empirical relation between the spin period and orbital period \citep{knigge_et_al_2011, enoto_et_al_2014}.
However, the spin period of 4U 1934+31 is among the longest known and the Corbet diagram offers little predictive ability other  
than requiring an orbit longer than a hundred days.  

The X-ray luminosity measures the mass accretion (\.{M}) onto the NS. Correcting to the 3295 pc distance, \.{M}$\,\approx\,$1$\times$10$^{16}$ g s$^{-1}$,
i.e. 2$\times$10$^{-10}$ M$_\odot$ yr$^{-1}$ \citep{enoto_et_al_2014}. Such a result may be used to constrain the orbital separation or, equivalently, the orbital period. Indeed, assuming 
a mass-loss rate for 4U~1954+31 of $\sim6\times10^{-7}$ M$_\sun$ yr$^{-1}$ \citep{beasor_et_al_2020}, a terminal wind velocity of 10 -- 30
km s$^{-1}$ \citep[Table 4 of][]{jura1990}, an orbital eccentricity 0.5, and an orbital separation $>5$ AU,
a rough estimation of the expected NS wind-accretion rate is  
$M_{acc}<3\times 10^{-9}$ M$_\sun$ yr$^{-1}$ \citep{bondi44}. In this context, the mass accretion rate onto the NS estimated from the observed X-ray luminosity implies  separation substantially larger than 5 AU.  As further evidence favoring this scenario, we note that the optical emission lines are 
not particularly strong. All together this requires a detached system with an orbital period much longer than 3 yrs. 

The semi-major axis of $\gtrsim$ 5 AU is the largest separation of any system 
classified as symbiotic. However, all other symbiotic systems involve low mass red giants and AGB stars.  For instance, the 
neutron star - M 3 III system V934 Her \citep{hinkle_et_al_2019}, has a similarly weak optical emission line
spectrum.  It has a 12 year orbital period and a 
semi-major axis of $\sim$3.7 AU.  
The neutron star - M6 III system
V2116 Oph \citep{hinkle_et_al_2006}, with a rich-emission line spectrum, has a semi-major axis of $\sim$1.6 AU. Classification
of 4U~1954+31, and other supergiant  -- degenerate mass transfer systems, in the symbiotic class needs review.

The spin period of 4U~1954+31 is variable with episodes of increasing and decreasing period.  Hence, a 
characteristic age cannot be found \citep{tauris_konar_2001}.
Models by \citet{ho_et_al_2020} show that, for long
spin periods, the period does not result from prolonged spin-down but is set during the initial $<$10$^{6}$ years of the NS life.  A 
very long spin period results from a 
combination of small mass accretion and large NS magnetic field \citep[equation 5 of][]{ho_et_al_2020}.  The current mass accretion rate and 
spin equilibrium period requires an unrealistically strong magnetic field. 
\citet{enoto_et_al_2014} find no evidence for an extreme magnetic field in the 4U~1954+31 NS and suggests a $\sim$10$^{13}$ G field.  
If the current M I was an unevolved B2 V star at the time of the SN, Reimer's law scales the mass loss down by a factor of 10$^3$.  If
the accretion similarly was 10$^3$ less, the magnetic field of the newly born NS was 10$^{15}$ G.  This 
suggests that the current M I was a MS star at the time of the SN.  Since most of the stellar life is spent in the core 
hydrogen burning stage, this seems a reasonable conclusion.

\subsection{OB Association Membership}\label{membership_section}

High energy X-ray sources are often associated
with star forming regions in the galactic arms \citep{chaty_2011}.  4U~1954+31 is located in Vulpecula, 3.7$^\circ$ from
the Vul OB2 association \citep{ruprecht_et_al_1982}.  This is a complex star forming region marking the
Orion-Cygnus arm. Gaia distances,
3295$^{+985}_{-62}$ pc  for 4U~1954+31 and $\sim$4.4 kpc for Vul OB2, place 4U~1954+31 in
the foreground of Vul OB2.  The proper motions of 4U~1954+31 and Vul OB2 are similar, but not identical
differing by $\sim$ 1.4 milliarcsecond in RA, i.e. 21 km s$^{-1}$ at 3.2 kpc.   We conclude that 4U~1954+31 is
likely not a member of Vul OB2.
On the other hand, the A and B supergiants in Vul OB2 \citep{turner_1980} exhibit a range of distance, proper motion, and
velocity.  The family of space motions is similar to the space motion of 4U~1954+31.
\citet{drazinos_et_al_2013} found that OB associations in
spiral galaxies have average diameters of $\sim$500 pc with star forming regions several times this size.
Association membership in this region is a complex problem and beyond the scope of this paper.
The general agreement of proper motions and velocities of the stars in this field, including 4U~1954+31,
does demonstrate that the 4U~1954+31 NS did not receive a large kick velocity from the SN.
 

\section{DISCUSSION}\label{discussion_section}

\subsection{Implications for the X-ray source}

The discovery that 4U~1954+31 is a late-type supergiant -- NS binary
impacts four parameters of interest in understanding the X-ray
properties.  First, the distance is $\gtrsim$ 1.9 times larger than
would be derived from the luminosity of the late-type star by assuming it is a giant. Hence, the luminosity of the
X-ray source is $\sim$4 times larger.  Second, the system is young
and the NS has evolved to its current state in a few tens of
megayears.  Third, the separation of the NS and M giant is greater
than previously seemed likely.  Fourth, the mass loss rate from the
late-type star is larger than previously believed.  Re-scaling the accretion rate onto the NS 
with the mass of the late-type star is a complex problem because a change of the M star mass
also affects the orbital parameters and, in turn, the accretion
rate.

The M supergiant mass-loss rate prescription of \citet{beasor_et_al_2020} and
the range of luminosity and mass in Table \ref{tab:system_parameters}
yield a broad possible range of  mass-loss rates, \.{M}, from 10$^{-5}$ to 10$^{-8}$ M$_\odot$ yr$^{-1}$. 
For the preferred values of 9 M$_\odot$ and
43880 L$_\odot$, \.{M}$\sim$6$\times$10$^{-7}$ M$_\odot$ yr$^{-1}$.  
A M4 giant has a
decidedly lower \.{M}, $\lesssim$10$^{-8}$ \citep{olofsson_et_al_2002,
groenewegen_et_al_2014}.  The mass loss process for M giants and supergiants
results from radiation pressure on dust.  The circumstellar flow
is cold, T$<$300 K, with the terminal wind velocity, v$_\infty$, in the range 10 to 30 km s${-1}$
\citep{jura1990}.
Studies of the bright M I $\alpha$ Ori are a source of detailed information.  
This star has a  terminal circumstellar expansion velocity of 14.3 km s$^{-1}$ \citep{huggins_1987}.  An extensive summary 
of observational parameters and references to the literature for $\alpha$ Ori can be found 
in \citet{dolan_et_al_2016}.  For $\alpha$ Ori \.{M}=2$\pm$1$\times$10$^{-6}$ M$_\odot$ yr$^{-1}$.  
By way of comparison, typical temperatures and terminal wind velocities for
the outflow in early-type HMXB are 30000 K and 1000 km s$^{-1}$.  Modified wind momentum  $ log(\dot{M} v_\infty R_{*}^{1/2}) $,
is commonly used to describe hot star winds.  Ignoring the very different mass loss physics 
between hot and cold winds, the modified wind momentum for $\alpha$ Ori is 27.7 (g cm s$^{-2}$ R$_\odot^{1/2}$).  
The modified wind momentum of the M supergiants 
falls on the lower end of the wind-momentum luminosity 
relation for massive stars \citep[see Figure 5 of][]{hainich_et_al_2020}.

In the case of wind accretion, the accretion rate onto the NS scales linearly with the mass donor mass-loss
rate, but decreases as v$_{wind}^{-3}$ \citep{boffin_jorissen_1988}
where v$_{wind}$=v$_\infty$+v$_{orb}$. More intense mass loss results in more
accretion, but a faster wind reduces the maximum radius below which
the wind is trapped into the gravitational potential well of the NS.   
For the long orbital 
period system 4U~1954+31, v$_{orb}$ is a few km s$^{-1}$, a fraction of the terminal wind velocity.  
As a first approximation this modulating term to the NS accretion can be ignored.
The X-ray luminosity of 4U 1954+31 implies a mass accretion rate of 
2x10$^{-10}$ M$_\odot$ yr$^{-1}$ \citep[][adjusted to 3.3 kpc distance]{enoto_et_al_2014}, 
implying that only a few parts per thousand of the mass loss from
the supergiant are captured by the NS.  This can be compared to the 
M 3 III -- NS system V934 Her. 
The wind velocity is likely similar but the mass loss rate is at least an 
order of magnitude less.  The mass accretion rate for the V934 Her NS  
is $\sim$10$^{-14}$ M$_\odot$ yr$^{-1}$ \citep{masetti_et_al_2002}.  This scales from that of 
4U~1954+31 with the difference in mass loss 
rates in agreement with the wind accretion relations of \citet{boffin_jorissen_1988}. 


\subsection{Evolution Scenario}

The standard scenario for the evolution of HMXB \citep{canal_et_al_1990,tauris_van_den_heuvel_2006, tauris_et_al_2017} 
invokes the evolution of   
two stars with zero-age MS masses sufficient for the stars to terminate in a core-collapse SN.
The binary system must have a sufficiently short period so that the stars will interact as
they evolve.  
Evolution of the more massive star results in  
large scale mass transfer through Roche lobe overflow.  Extreme mass loss from 
this star produces a roughly 3.5 M$_\odot$ helium star,
i.e. the stripped core of the massive star.  
The helium star undergoes a low luminosity core collapse SN resulting in a system with a NS and a massive 
star.  Mass accretion onto the secondary widens the orbit resulting in an eccentric orbit of period around 15 years.    
The system now appears as a HMXB as mass loss from the former secondary, now the high mass primary, is accreted onto the NS.
As the massive star evolves mass loss decreases the orbital period.  The terminal stage of the HMXB 
is a common envelope (CE) phase that ejects mass from the system  
producing a helium star.
Following the core collapse SN of the helium star a NS - NS binary results with an orbital period of hours.  
A system of this type radiates gravity wave radiation and, if the orbit is short enough, will merge.

The 4U~1954+31 system originated with two high-mass stars.  The lack of a kick velocity for 4U~1954+31 supports an 
origin of the NS from an ultra-stripped progenitor, either through an electron capture instability SN \citep{kochanek_et_al_2019}
or low-mass iron core-collapse SN \citep{tauris_et_al_2017}.  In addition, 
the current orbit is both eccentric and has a period of multiple years.

\subsection{Mass limits and the super-AGB}

The FuNS models shown in Figure \ref{fig:hrd} follow the evolutionary path up to the final stages of burning.
On the base of the observed location in the HR diagram, a range of mass, 
perhaps as low as 7\,$M_{\sun}$ and as high as 15\,$M_{\sun}$, must be considered for the M star in
4U~1954+31.  
Models with $M\geq 10$ M$_\odot$ attain the brightest point of the corresponding track during
the C burning and, later on, their luminosity does not change for a few thousand years 
until the final collapse. 
The two magenta lines in Figure \ref{fig:hrd} are for the two smaller masses, namely 7 and 9\,$M_{\sun}$. 
The smaller mass coincides with the minimum mass of a star that can attain the 
conditions for the C ignition. 
Less massive stars skip the C burning, enter the AGB and terminate their life as CO white dwarfs. 
The 7\,$M_{\sun}$ star, ignites C off centre in degenerate conditions and, after an incomplete C burning,
 enters the super-AGB phase. 
Thus, the brightest point of the track corresponds to the luminosity at the beginning of the super-AGB. 
This star will end as a C-O-Ne WD of mass of about 1.1\,$M_{\sun}$. 

The 9\,$M_{\sun}$ model, on the contrary, experiences an almost complete C-burning phase that
leaves a degenerate O-Ne-Mg core.   
Then, this star also moves to the super-AGB phase. 
The brightest point of the track is attained after the first few thermal pulses. 
The C ignition occurs at log$(L/L_{\sun})\,\approx\,$4.5, but in the 
following evolution, the luminosity increases up to log$(L/L_{\sun})$~$\sim$~5.1. 
This transition at high luminosity is quite fast. 
During the super-AGB the luminosity may further increase
because of the occurrence of the hot bottom burning \citep{doherty_et_al_2014}. 
The duration of this phase, depending on the mass loss rate, 
could be a few times 10$^4$ years (9\,$M_{\sun}$), up to 10$^5$ years \citep[7\,$M_\odot$, see][]{doherty_et_al_2015}. 
Since in the 9\,$M_{\sun}$ model the degenerate core is about 1.3\,$M_{\sun}$, very close to the 
Chandrasekhar limit, it is possible that the core will undergo an electron capture instability, giving rise to 
either a core-collapse or a thermonuclear explosion \citep[e-capture SN,][]{woosley_heger_2015}.

The best match of the observed HRD location is obtained for an initial mass between 
7 and 11\,$M_{\sun}$. 
In general, we have to consider the time spent by the star in each portion of the track. 
In this context, a more massive object cannot be excluded, even if its effective temperature 
appears higher than observed. 
Indeed, the effective temperature of a model may be affected by radiative opacity 
uncertainties (atomic and molecular opacity, in particular). 
In addition, it also depends on the adopted mixing length parameter that, as usual, 
has been calibrated using the Sun.  It is not clear whether this calibration is appropriate for a red supergiant. 
A further phenomenon that may affect the quoted mass range is rotation. Indeed, a fast rotation
during the main sequence phase implies larger final core masses, and, in turn, higher 
luminosities during the post-MS evolution.
Unfortunately, the MS rotation velocity of the M star in 4U~1954+31 is indeterminate.   
In spite of these uncertainties, we feel confident excluding an initial mass $<$7\,$M_{\sun}$. 

\subsection{Comparison with Abundances}

The surface abundances of the red supergiant in the 4U~1954+31 system are the result of the 
original composition, stellar evolution and mixing processes and, possibly,  mass transfer from the proto-NS primary as it evolved.
Single star models of 9-15\,$M_{\sun}$, starting from solar abundance ratios, 
indicate that [C/Fe], [N/Fe], and [O/Fe] after the first dredge up (FDU) should be $-$0.25, +0.55, and $-$0.07, respectively. 
The $^{12}$C/$^{13}$C ratio is predicted to be $\sim$20.  $^{16}$O/$^{17}$O and $^{16}$O/$^{18}$O are 
both around 750 in the 13\,$M_{\sun}$ model and 650-700 in the 9\,$M_{\sun}$ model. 
In a 10.8 M$_\odot$ model by \citet{takahashi_et_al_2013}, the post-FDU
composition is $^{12}$C/$^{13}$C=20, $^{16}$O/$^{17}$O=810, and $^{16}$O/$^{18}$O=660.  
Takahashi et al. report of a dredge-out episode occurring at the end of the 
C-burning phase, causing a sharp increase of the $^{18}$O abundance at the stellar surface 
and, in turn, a substantial drop of $^{16}$O/$^{18}$O to about 40.
In the case of a super-AGB star, the HBB should also produce sizeable modifications of the post-FDU composition.
A significant increase in the N abundance is expected, while $^{12}$C should 
be reduced \citep{doherty_et_al_2014}. 
Isotopic ratios of C and O are predicted to change as well: 
$^{12}$C/$^{13}$C and $^{16}$O/$^{17}$O are both expected to go 
down, while $^{16}$O/$^{18}$O should increase. Then, the possible coupling of HBB and third dredge up may cause an increase of the C+N+O.

Comparison of these predictions with the derived 4U~1954+31 abundances should constrain its evolutionary status.
The predicted post-FDU nitrogen is in good agreement with the observed value, while the predicted C and O depletion are 
slightly smaller than observed, but within the uncertainty.  In addition, the CNO abundances sum up, 
within the uncertainties, to the solar CNO value scaled to the observed metallicity, in accordance with 
measurements from other M supergiants \citep{lyubimkov_et_al_2019}. In contrast,  no clear evidence of HBB or third dredge-up is found.  
This, however, does not exclude that the M supergiant in 4U~1954+31 could be a star with mass 
in the range 7-10  M$_\odot$, provided that its present evolutionary status is beyond the C ignition 
and before the occurrence  of the first thermal pulses during the super-AGB phase. 

At variance with the elemental CNO abundances, the post-FDU isotopic composition does not comply with the observed values. 
The measured $^{12}$C/$^{13}$C, 9$\pm$2, is clearly less than the predicted value. 
\citet{harris_lambert_1984} found similar $^{12}$C/$^{13}$C for the supergiants $\alpha$ Ori and $\alpha$ Sco.
The oxygen isotopic ratios come with considerable observational uncertainties, in 
particular for $^{16}$O/$^{18}$O where the measured value relies on a single line.
Both the observed $^{16}$O/$^{17}$O and $^{16}$O/$^{18}$O ratios appear higher than the values expected after the FDU. In the case of a 
super-AGB star undergoing HBB,  the match with the observed $^{16}$O/$^{18}$O would be better but the difference with the observed $^{16}$O/$^{17}$O would be even worse. 
Note, however, that our isotopic predictions start from solar abundance ratios and that a 
deviation from this hypothesis is not unusual in young star forming regions influenced by supernova explosions. 
In any case, due to the large uncertainties, we cannot 
put much weight on the comparison of observed and modeled oxygen isotopic ratios.

Mass transfer from the more evolved star may be a significant factor in altering the abundance pattern of a star belonging to a close binary system.
The lack of kick velocity 
for the NS in 4U~1954+31 suggests that the MS primary became a He star before core collapse.  The mass
of this star at the time of core collapse was likely in 
the 3 - 4 M$_\odot$ range \citep{tauris_et_al_2017}.  Since this star  
had a main sequence mass $\gtrsim$ 9 M$_\odot$ it lost $\gtrsim$5 M$_\odot$.  The mass 
loss process was most likely Roche lobe overflow onto the secondary \citep{tauris_et_al_2017}.  The evolution of 
a system of this type is complex and depends on the orbit.  The Roche lobe mass 
transfer will spin up the accretor, stopping mass accretion \citep{de_mink_et_al_2013}.  For a 
Roche lobe filling supergiant a common envelope develops, since as the supergiant loses mass it continues to expand.  
The expectation is that the envelope of the donor is ejected 
and the secondary star left relatively unaffected \citep{de_mink_et_al_2013}.  

In addition to this mass transfer event, the present-day supergiant then was exposed to the SN ejecta.
Either accretion of the material ejected by the SN or ablation of the stellar envelope are possible. 
Based on the measured abundances, the only indication of a possible
SN pollution are the over-abundances of the r-process element Sc and the iron peak element Mn.  
However, the Sc abundance is suspect.  \citet{thorsbro_et_al_2018} found that for the near-infrared Sc lines non-LTE effects result in the 
abundance being overestimated at T$_{eff}$ $<$ 3800\,K. 
Pollution may differ from one element to another, depending on the initial abundance. 
In addition, a large overabundance of $\alpha$-elements, such as O or Ti, are expected 
in case of a SN pollution, in contrast with the observed abundance pattern. 
Alternatively, it is also possible that the fast wind from the SN had stripped part 
of the envelope, leaving no trace of either the SN or of mass transferred before the SN. 

\subsection{Ages and Life Expectancy}\label{ages_section}

The creation of the NS erases most information about the progenitor. Certainly the progenitor was massive enough
to undergo a core collapse, but not too massive, otherwise the compact remnant would be a BH instead of a NS.  \citet{hills_1983}
discusses the effects of the sudden mass loss, i.e. a SN, on the orbit.  The survival of the binary, in the event that the SN was from a massive
star core collapse, potentially provides limits on the mass.  However, 
the lack of a kick velocity requires that the SN resulted from an ultra-stripped star.  
The progenitor of the NS, since it evolved faster, was initially the more massive star in the binary.  Since the M I possibly could evolve
into a massive WD, the mass limits of the MS primary are just those of a star 
producing a NS, 9 M$_\odot$~$\lesssim$~M~$\lesssim$~25 M$_\odot$. The upper mass bound actually depends on the upper mass limit for a NS, the so-called Tolman-Oppenheimer-Volkoff mass, a quantity affected by large uncertainties. In addition, in the case of a rather small progenitor, the collapse may form a BH rather than a NS, directly or by fall back, depending on the compactness of the pre-supernova structure \citep{oconnor2011}. The
lifetime of a 25 M$_\odot$ is 7 Myr and of a 9 M$_\odot$ 30 Myr.  The M I star 
has a mass 7 M$_\odot$~$\lesssim$~M~$\lesssim$~15 M$_\odot$.  Since
this star has exhausted core hydrogen it is 
effectively at the end of its life and its age is 
50 Myr (7 M$_\odot$) to 12 Myr (15 M$_\odot$). The lack of a SN remnant constrains the time that has elapsed since 
the SN to $\gtrsim$10$^5$ years \citep{stafford_et_al_2019}, giving the minimum age of the NS. 
The difference in stellar lifetimes sets the maximum age of the NS at 43 Myr.  

There are $\sim$114 HMXB known \citep{chaty_2015}.  These systems all contain hot stars with 4U~1954+31 the only late-type system. This likely reflects the lifetime 
as a red rather than blue supergiant.   In the mass range 10 -- 20 M$_\odot$, the ``normal'' stellar wind is 
not enough to completely erode the H-rich envelope and the star remains in the red supergiant branch up to the core collapse. 
In the case of enhanced mass loss, for instance in close binaries, the star can return to the blue before the final 
collapse. In the mass range 7 -- 10 M$_\odot$, the situation can be more complicated for the stars that enter a super-AGB phase 
after C burning. During the super-AGB, the core mass increases, while the envelope is progressively lost. If the core 
mass attains the Chandrasekhar limit before the envelope is reduced down to a few tenths of a solar mass, a SN occurs when the star 
is a red supergiant. If not, i.e., if the H-rich envelope is almost completely eroded, the star moves to the blue, ending as a (C-)O-Ne-Mg WD. 

\section{CONCLUSIONS}\label{conclusion_section}

4U~1954+31 is shown to be a M supergiant -- NS binary.  The M companion to the NS had previously been identified as a 
M III \citep{masetti_et_al_2006}.  Gaia data show that the luminosity of the M star 
is too high for the star to be a normal giant.  The near-IR spectrum is a good match to spectral type M4 I.  From the 
excitation temperature of the 1.6 $\mu$m CO,
$T_{\rm{eff}}$ = 3450 $^{+100}_{-50}$ K, in agreement with the spectral type.  Optical 
through mid-IR photometry combined with the Gaia parallax give a  
luminosity = 43880$^{+34070}_{-15900}$  L$_\odot$ and radius = 586 $^{+188}_{-127}$ R$_\odot$.  Spectrum 
synthesis suggests a surface gravity (log\,g) of  -0.15$\pm$0.25  (cm s$^{-1}$) corresponding to a 
9 -- 12 M$_\odot$ star.  Evolutionary tracks indicate that the mass is 9 $_{-2}^{+6}$ M$_\odot$.  A time series of 
velocities from high resolution optical spectra tracks part of the orbit.  The orbital period is $\gtrsim$3 yrs, 
in agreement with the limit set from the Roche lobe of a 9 M$_\odot$ supergiant plus a 1.4 M$_\odot$ NS binary.

Although the existence of a binary system containing a M supergiant plus NS is expected from stellar evolution theory,
this is the first such system to be identified.  The binary shares characteristics of both high and low 
mass X-ray binaries.  It also has been grouped with the symbiotic X-ray binaries.  However, no other symbiotic binary
contains a supergiant and the standard model of the symbiotic class excludes it \citep{kenyon_webbink_1984}.  Considered as a 
symbiotic system, it has the largest known semi-major axis, $>$ 4.5 AU.  

The lack of a 
kick velocity argues for the origin of the NS in a low luminosity core collapse SN.  This requires that 
the mass of the primary to be reduced by at least 5 M$_\odot$ before core collapse, likely by a CE event.
The current M supergiant was a B MS star when the originally more massive star went SN.  The 
abundances for the M I are not exceptional.  No trace 
of the CE or SN events appears in 
the abundances.  This is either the result of the SN ablating the surface of the B MS star or subsequent 
mixing of the surface material into the envelope as the B star evolved. Roche lobe mass transfer onto the MS B star
could have spun up the star, enhancing meridional mixing.  Using the lowest possible mass 
of the M supergiant to constrain the age of the 
4U~1954+31 NS, the NS age is $\lesssim$43 Myr.  

\acknowledgments 
The authors thank Bernhard Aringer for giving access to his spectral synthesis code and 
for support with the computations.  NOAO reference librarian Sharon Hunt
helped us access obscure reference materials.  SM plot, developed by Robert Lupton and Patricia Monger, was used in the 
production of some figures.
This research was facilitated by the SIMBAD database, operated by CDS in Strasbourg, France, the VizieR catalogue access tool,
CDS, Strasbourg, France (DOI: 10.26093/cds/vizier),
and the Astrophysics Data System Abstract Service, operated by the Smithsonian 
Astrophysical Observatory under NASA Cooperative Agreement NNX16AC86A. This work made use of data from the 
European Space Agency (ESA) mission $Gaia$ (\url{https://www.cosmos.esa.int/gaia}), processed 
by the Gaia Data Processing and Analysis Consortium (DPAC).  
Funding for the DPAC has been provided by national institutions, in particular the institutions 
participating in the $Gaia$ Multilateral Agreement.  This work used IGRINS,
developed under a collaboration between the University of Texas at Austin and the Korea Astronomy and Space 
Science Institute (KASI) with the financial support of the US National Science Foundation under 
grants AST-1229522 and AST-1702267, of the University of Texas at Austin, and of the Korean GMT Project of KASI. 
This publication makes use of data obtained with the Lowell Discovery Telescope (LDT) at Lowell Observatory. Lowell is a 
private, non-profit institution dedicated to astrophysical research and public appreciation of astronomy 
and operates the LDT in partnership with Boston University, the University of Maryland, the 
University of Toledo, Northern Arizona University and Yale University.
NSF's National Optical-Infrared Astronomy Research Laboratory (NOIRLab), which supersedes 
the National Optical Astronomy Observatory (NOAO), is operated by the Association of Universities for 
Research in Astronomy under a cooperative agreement with the National Science Foundation. 
IRAF (Image Reduction and Analysis Facility) software was distributed by NOAO.  KH and RJ express 
their thanks to the NOAO Office of Science and the NOIRLab RSS group for support of this research.  
The research at Tennessee State University was supported in part by the 
State of Tennessee through its Centers of Excellence program.

OCHID identification numbers:~~KENNETH H. HINKLE [0000-0002-2726-4247], THOMAS LEBZELTER [0000-0002-0702-7551],
FRANCIS C. FEKEL [0000-0002-9413-3896], OSCAR STRANIERO [0000-0002-5514-6125],
RICHARD R. JOYCE [0000-0003-0201-5241], LISA PRATO [0000-0001-7998-226X],
NICOLE KARNATH [0000-0003-3682-854X], NOLAN HABEL [0000-0002-2667-1676]

\clearpage

\begin{deluxetable}{lcl}
\tablewidth{0pt}
\tablecaption{Table of Observations and Velocities} \label{tab:observations}
    \tablehead{\colhead{HJD} & \colhead{RV\tablenotemark{a}} & \colhead{Observatory/} \\
\colhead{-2450000} & \colhead{(km s$^{-1}$)} & \colhead{Spectrograph}
}
\startdata
8039.711  & -1.4 &  AST/FFES \\    
8041.772  & -1.2 &  AST/FFES  \\
8042.632  & -1.4 &  AST/FFES \\
8043.624  & -1.4 &  AST/FFES \\
8226.859  & 0.1 & AST/FFES \\  
8417.708  & 3.3 & AST/FFES \\
8423.686  & 3.5 & AST/FFES  \\
8439.661  & 4.1 & AST/FFES  \\
8445.616  & 4.0 & LDT/IGRINS \\
8456.601  & 5.1 & AST/FFES  \\
8465.621  & 5.4 & AST/FFES \\
8532.005  & 6.1 & AST/FFES  \\
8540.965  & 6.0 & AST/FFES \\
8565.999  & 6.1 & AST/FFES   \\
8599.945  & 5.8 & AST/FFES   \\
8627.932  & 6.3 & AST/FFES \\
8661.804  & 6.0 & AST/FFES \\
8739.754  & 6.4 & AST/FFES  \\
8767.642  & 5.9 & AST/FFES \\
8768.642  & 6.2 & AST/FFES \\
8800.620  & 6.4 & AST/FFES  \\
8801.620  & 6.4 & AST/FFES  \\
8963.837  & 5.3 & AST/FFES  \\
8964.920  & 5.8 & AST/FFES   \\
8965.921  & 5.7 & AST/FFES  \\  
9020.692 & 6.1 & AST/FFES \\
9024.794 & 6.1 & AST/FFES \\
9025.794 & 6.0 & AST/FFES \\
\enddata
\tablenotetext{a}{Heliocentric. To correct to LSR add 17.77 km s$^{-1}$}
\end{deluxetable}

\clearpage

\begin{deluxetable}{lcl}
\tablewidth{0pt}
\tablecaption{Parameters of the 4U~1954+31 M supergiant} \label{tab:system_parameters}
\tablehead{\colhead{Parameter} & \colhead{Value} & \colhead{Source}
}
\startdata
Distance              & 3295$^{+985}_{-631}$ pc             & $Gaia$ \citet{bailer-jones_et_al_2018}   \\
Spec Type             & M4 I                         & Fig. 1; T$_{eff}$ \& luminosity   \\
$T_{\rm{eff}}$        & 3450 $^{+100}_{-50}$ K          & Sp.Ty.; CO T$_{exc}$            \\
Luminosity            & 43880$^{+34070}_{-15900}$  L$_\odot$    & Photometry, see text; Fig. 2    \\
Radius                &  586 $^{+188}_{-127}$ R$_\odot$        & Fig. 2 \\
Mass                  &  9$^{+6}_{-2}$ M$_\odot$            & Luminosity \& Evol. tracks    \\ 
Surface gravity (log\,g) &  -0.15$\pm$0.25  (cm s$^{-1}$) & Mass \& radius       \\
$[Fe/H]$              &  -0.4 $\pm$ 0.2             & See text      \\
Age                   &   $\sim$12 - 50 Myr           & Evol. tracks          \\
\enddata
\end{deluxetable}

\clearpage

\begin{deluxetable}{lr|lr}
\tablecaption{Atomic Line List\label{t:usedatomiclines}}
\tablehead{
\colhead{Element} & \colhead{wavenumber} & \colhead{Element} & \colhead{wavenumber}\\
\colhead{}        & \colhead{(cm$^{-1}$)}  & \colhead{}      & \colhead{(cm$^{-1}$)}
}
\startdata
\ion{Na}{1} & 4527.0 & \ion{Fe}{1} & 4732.7\\ 
            & 4532.6 &             & 5786.3\\
\ion{Al}{1} & 4739.6 &             & 5786.3\\
            & 5963.8 &             & 5810.9\\
            & 5968.3 &             & 5812.3\\
            & 5979.6 &             & 5814.8\\
\ion{Si}{1} & 5593.3 &             & 5825.5\\
            & 5940.8 &             & 5867.7\\
            & 6165.2 &             & 5876.9\\
            & 6185.0 & \ion{V}{1}  & 4582.4\\
            & 6211.5 &             & 4786.5\\
            & 6224.9 &             & 6033.2\\
            & 6313.9 &             & 6277.8\\
\ion{Sc}{1} & 4489.8 & \ion{Mn}{1} & 6569.5\\
            & 4501.8 & \ion{Ti}{1} & 4419.4\\
            & 4530.8 &             & 4454.3\\
            & 4533.5 &             & 4481.0\\
            & 4583.3 &             & 4488.3\\
            & 4600.6 &             & 4496.6\\
\ion{Fe}{1} & 4382.3 &             & 4501.0\\
            & 4419.7 &             & 4543.3\\
            & 4444.4 &             & 4565.5\\
            & 4458.0 &             & 4589.5\\
            & 4468.0 &             & 4691.7\\ 
            & 4507.7 &             & 4793.6\\
            & 4654.2 &             & 5874.5\\ 
            & 4707.1 &             & 5883.3\\
            & 4712.1 &             & 6431.7\\ 
            & 4720.6 & & \\   
\enddata
\end{deluxetable}

\clearpage

\startlongtable
\begin{deluxetable}{lcr}
\tablecaption{Molecular Line List\label{t:usedmollines}}
\tablehead{
\colhead{Molecule} & \colhead{transition} & \colhead{wavenumber}\\
\colhead{}         & \colhead{}           & \colhead{(cm$^{-1}$)}
}
\startdata
$^{12}$CO & 9-6 head & 5920--5940\\
          & 8-5 head & 6005--6020\\
          & 7-4 head & 6086--6098\\
          & 6-3 head & 6167--6177\\
          & 5-2 head & 6245--6257\\
          & 4-2 R18 & 4212.7\\
          & 2-0 P11 & 4213.9\\
          & 2-0 P9 & 4223.0\\
          & 3-1 R4 & 4225.2\\
          & 4-2 R24 & 4225.7\\
          & 4-2 R27 & 4231.2\\
          & 4-2 R28 & 4232.8\\
          & 3-1 R7 & 4235.1\\
          & 4-2 R31 & 4237.4\\
          & 3-1 R8 & 4238.3\\
          & 3-1 R9 & 4241.4\\
          & 4-2 R36 & 4243.6\\
          & 3-1 R84 & 4260.2\\
          & 2-0 R1 & 4267.5\\
          & 6-3 P18 & 6030.1\\
          & 5-2 P30 & 6033.9\\
          & 6-3 P17 & 6035.6\\
          & 4-1 P35 & 6076.4\\
          & 7-4 R46 & 6087.9\\
          & 7-4 R20 & 6088.0\\
          & 4-1 P28 & 6125.4\\
          & 6-3 R63 & 6127.9\\
          & 6-3 R16 & 6160.7\\
          & 6-3 R17 & 6162.6\\
          & 4-1 P19 & 6181.0\\
          & 4-1 P18 & 6186.7\\
          & 5-2 R69 & 6187.1\\
          & 4-1 P16 & 6197.7\\
          & 5-2 R64 & 6205.6\\
          & 5-2 R61 & 6215.3\\
          & 5-2 R6 & 6215.6\\
          & 4-1 R11 & 6308.7\\
          & 4-1 R19 & 6325.2\\
          & 4-1 R48 & 6326.4\\
          & 4-1 R20 & 6326.7\\
          & 3-0 R8 & 6380.3\\
          & 3-0 R53 & 6398.3\\
$^{13}$CO & 4-2 head & \\
          & 3-1 head & \\
          & 2-0 head & \\
          & 3-1 R33 & 4200.3\\
          & 3-1 R69 & 4201.0\\
          & 2-0 R19 & 4226.4\\
          & 2-0 R21 & 4230.9\\
          & 2-0 R24 & 4237.1\\
          & 2-0 R35 & 4254.5\\
          & 2-0 R37 & 4256.8\\
          & 2-0 R38 & 4257.8\\
          & 2-0 R40 & 4259.7\\
          & 2-0 R41 & 4260.5\\
$^{12}$CN & 1-3 P$_2$ 38.5 & 4499.2 \\
          & 0-2 Q$_2$ 57.5 & 4555.0\\
          & 1-3 P$_2$ 32.5 & 4584.7\\
          & 1-3 P$_1$ 31.5 & 4596.3\\
          & 0-2 Q$_1$ 55.5 & 4596.9\\
          & 1-3 P$_2$ 31.5 & 4598.0\\
          & 1-3 Q$_1$ 40.5 & 4606.5\\
          & 1-3 P$_2$ 30.5 & 4611.0\\
          & 0-2 Q$_1$ 54.5 & 4613.1\\
          & 1-3 P$_1$ 29.5 & 4620.7\\
          & 0-2 Q$_2$ 53.5 & 4621.8\\
          & 1-3 P$_2$ 29.5 & 4623.7\\
          & 0-2 Q$_2$ 52.5 & 4637.8\\
          & 0-2 Q$_2$ 50.5 & 4668.9\\
          & 1-3 P$_2$ 25.5 & 4671.8\\
          & 1-3 Q$_1$ 34.5 & 4672.2\\
          & 1-3 P$_2$ 24.5 & 4683.2\\
          & 0-2 Q$_2$ 49.5 & 4684.0\\
          & 1-3 Q$_2$ 33.5 & 4685.2\\
          & 1-3 P$_1$ 23.5 & 4686.4\\
          & 0-2 Q$_1$ 49.5 & 4689.3\\
          & 1-3 Q$_2$ 32.5 & 4695.4\\
          & 1-3 P$_1$ 22.5 & 4696.2\\
          & 0-2 Q$_2$ 48.5 & 4698.8\\
          & 1-3 Q$_1$ 31.5 & 4701.0\\
          & 0-1 P$_1$ 58.5 & 6308.4\\
          & 0-1 Q$_2$ 66.5 & 6324.4\\
          & 1-2 P$_2$ 42.5 & 6393.1\\
          & 0-1 P$_1$ 52.5 & 6437.1\\
          & 1-2 P$_2$ 39.5 & 6445.0\\
          & 0-1 Q$_2$ 60.5 & 6451.8\\
          & 0-1 P$_1$ 51.5 & 6457.4\\
          & 1-2 P$_2$ 37.5 & 6477.9\\
OH & 4-2 P$_{2f}$ 5.5 & 6080.0\\
   & 2-0 P$_{2f}$ 15.5 & 6079.6\\
   & 3-1 P$_{1f}$ 11.5 & 6107.8\\
   & 3-1 P$_{2e}$ 10.5 & 6112.8\\
   & 3-1 P$_{2f}$ 10.5 & 6113.7\\
   & 2-0 P$_{1f}$ 14.5 & 6219.5\\
   & 2-0 P$_{2e}$ 13.5 & 6223.0\\
   & 2-0 P$_{2f}$ 13.5 & 6224.3\\
   & 3-1 P$_{1e}$ 9.5 & 6227.8\\
   & 3-1 P$_{1e}$ 8.5 & 6283.5\\
   & 3-1 P$_{2e}$ 6.5 & 6344.8\\
   & 3.1 P$_{2f}$ 6.5 & 6345.2\\
   & 3-1 P$_{2e}$ 5.5 & 6397.3\\
   & 3-1 P$_{2f}$ 5.5 & 6397.6\\
   & 2-0 P$_{1e}$ 11.5 & 6421.4\\
   & 2-0 P$_{1f}$ 10.5 & 6482.3\\
   & 2-0 P$_{1e}$ 10.5 & 6483.5\\
   & 2-0 P$_{2e}$ 9.5 & 6487.9\\
   & 2-0 P$_{2f}$ 9.5 & 6488.7\\
   & 3-1 P$_{1f}$ 4.5 & 6479.8\\
HF & 1-0 R3 & 4279.9\\
\enddata
\end{deluxetable}

\clearpage

\begin{deluxetable}{lcrr}
\tablecaption{Elemental abundances of the M supergiant\label{tab:abundances}}
\tablewidth{0pt}
\tablehead{
\colhead{Element} & \colhead{Number of} & \colhead{Abundance} & \colhead{Uncertainty}\\
\colhead{ } & \colhead{lines} & \colhead{relative to solar} & \colhead{}
}
\startdata
C & 37\tablenotemark{a} & -0.5 & $\pm$0.1 \\ 
N & 33 & +0.1 & $\pm$0.15 \\ 
O & 20 & -0.4 & $\pm$0.15 \\
F & 1 & -0.9 & $\pm$0.1 \\
Na & 2 & -0.1 & $\pm$0.1 \\
Al & 4 & -0.2 & $\pm$0.1 \\
Si & 7 & -0.5 & $\pm$0.2 \\
Sc & 6 & +0.2 & $\pm$0.05 \\
Ti & 14 & -0.25 & $\pm$0.1 \\
V & 4 & -0.15 & $\pm$0.2 \\
Mn & 1 & +0.3 & $\pm$0.1 \\
Fe & 19 & -0.4 & $\pm$0.2 \\
\enddata
\tablenotetext{a}{Number of $^{12}$CO lines used for the analysis. In addition, several band heads were included in the fit.}
\end{deluxetable}

\clearpage

\end{document}